\documentclass[aps,pra,reprint]{revtex4-2}

\usepackage[british]{babel}
\usepackage{graphicx}
\usepackage{xcolor}
\usepackage{amssymb}
\usepackage{amsmath}
\usepackage{tabularx}
\usepackage{multirow}
\usepackage{booktabs}

\usepackage{placeins}
\usepackage{nicefrac}

\usepackage[normalem]{ulem}
\usepackage{color}

\newcommand{\ket}[1]{|{#1}\rangle}

\begin{document}

%
%
\title{Experimental anonymous conference key agreement using linear cluster states}
	
\author{Lukas Rückle$^{1,2}$, Jakob Budde$^{1,2}$, Jarn de Jong$^{3}$, Frederik Hahn$^{3,4}$, Anna Pappa$^{3,5}$, Stefanie Barz$^{1,2}$}
\affiliation{
$^1$Institute for Functional Matter and Quantum Technologies, Universit{\"a}t Stuttgart, 70569 Stuttgart, Germany \\
$^2$Center for Integrated Quantum Science and Technology (IQST), Universit{\"a}t Stuttgart, 70569 Stuttgart, Germany\\
$^3$Electrical Engineering and Computer Science Department, Technische Universit{\"a}t Berlin, 10587 Berlin, Germany\\
$^4$Dahlem Center for Complex Quantum Systems, Freie Universit{\"a}t Berlin, 14195 Berlin, Germany\\
$^5$Fraunhofer-Institute for Open Communication Systems FOKUS,  10589 Berlin, Germany \looseness=-1}

%
%
\begin{abstract}
Multipartite entanglement enables secure and anonymous key exchange between multiple parties in a network. In particular Greenberger--Horne--Zeilinger (GHZ) states have been introduced as resource states for anonymous key exchange protocols, in which an anonymous subset of parties within a larger network establishes a secret key. However, the use of other types of multipartite entanglement for such protocols remains relatively unexplored. Here we demonstrate that linear cluster states can serve as a versatile and potentially scalable resource in such applications.
We implemented an anonymous key exchange protocol with four photons in a linear cluster state and established a shared key between three parties in our network.
We show how to optimize the protocol parameters to account for noise and to maximize the finite key rate under realistic conditions. As cluster states have been established as a flexible resource in quantum computation, we expect that our demonstration provides a first step towards their hybrid use for networked computing and communication.
\end{abstract}

\maketitle

%
%
\section{Introduction}
Quantum communication has been expanded from the initially proposed bi-partite key exchange~\cite{bennett1984quantum, ekert1991quantum} to networked settings~\cite{peev2009secoqc,sasaki2011field,dynes2019cambridge,wengerowsky2018entanglement}.
One particularly interesting application of quantum networks is conference key agreement~\cite{grasselli2018finite,murta2020quantum}.
In such protocols, multipartite entangled states are used to realize a key exchange in a quantum network. 
It has been shown that such a networked key exchange is possible by sharing Greenberger--Horne--Zeilinger (GHZ) states in a network~\cite{chen2004multi}. Their quantum correlations can be harnessed for establishing a joint key and for performing verification. In the latter step, an eavesdropper or any other deviation in the protocol can be detected, similar to the bi-partite case, making the protocol \textit{secure}~\cite{grasselli2018finite}.

Furthermore, the use of multipartite entanglement made it possible to efficiently realize another security feature beyond security: \textit{anonymity}.
By exploiting the particular properties of GHZ states, multiple parties can communicate in a quantum network with their identities protected~\cite{hahn2020}.
In other words, a key is exchanged between a subset of parties of a network while it remains hidden which parties belong to this subset. 
Such anonymous quantum conference key agreement also allows verification and thus the detection of deviating parties or eavesdroppers.
There are various implementations of conference key agreement and its anonymous equivalent in quantum networks ~\cite{proietti_experimental_2020, thalacker2021anonymous,pickston2022experimental}. 

So far, many protocols for conference key agreement build strongly on the particular correlations of GHZ states. 
This raises the question whether multipartite entangled states other than GHZ states are a suitable resource for such protocols.
This invites the question whether multipartite entangled states other than GHZ might be suitable as a resource for such protocols.
Of particular interest with regard to scalability are physical quantum networks that, due to their topology or physical hardware, favour building up links in the form of linear cluster states.

Here, we study the use of linear cluster states as a resource for anonymous conference key agreement between a subset of parties in a larger network. We generate four-photon cluster states and demonstrate that they provide a basis for key exchange between three parties of our network, by implementing a recently introduced protocol~\cite{de2022anonymous}.
Specifically, we exchange a key with a length of about 40\,kbit and demonstrate the encryption, sharing and decryption of a picture over the network. We evaluate the success rates of the protocol for different network configurations and examine the influence of experimental imperfections, in particular how the parameters of the protocol can be adapted to certain noise values.

While we are focusing on a key exchange in a network of four parties, the protocol can be scaled to anonymous three-partite communication in a larger network.
As such, our work establishes the potential of cluster states beyond applications in quantum computing~\cite{raussendorf2001one}.

%
%

\section{Protocol}\label{sec:theory}

We start by introducing the main steps of the protocol~\cite{de2022anonymous}.
The first step is the creation of the resource state, a linear cluster state, followed by its distribution to all parties in the network. 
Such a linear cluster state can, in general, be created by each party holding a qubit in the state $\ket{+}=\nicefrac{1}{\sqrt{2}}(\ket{0}+\ket{1})$, where $\ket{0}$ and $\ket{1}$ are the computational basis states. The qubits are then entangled by applying \textit{CPhase} gates between pairs of neighbouring qubits~\cite{hein2006entanglement}. Here, the action of the CPhase gate is $\mbox{CPhase}\ket{ij}=(-1)^{ij}\ket{ij}$~\cite{nielsen2002quantum}. 


\begin{figure}[t!]
	\begin{centering}
		\includegraphics[width=.47\textwidth]{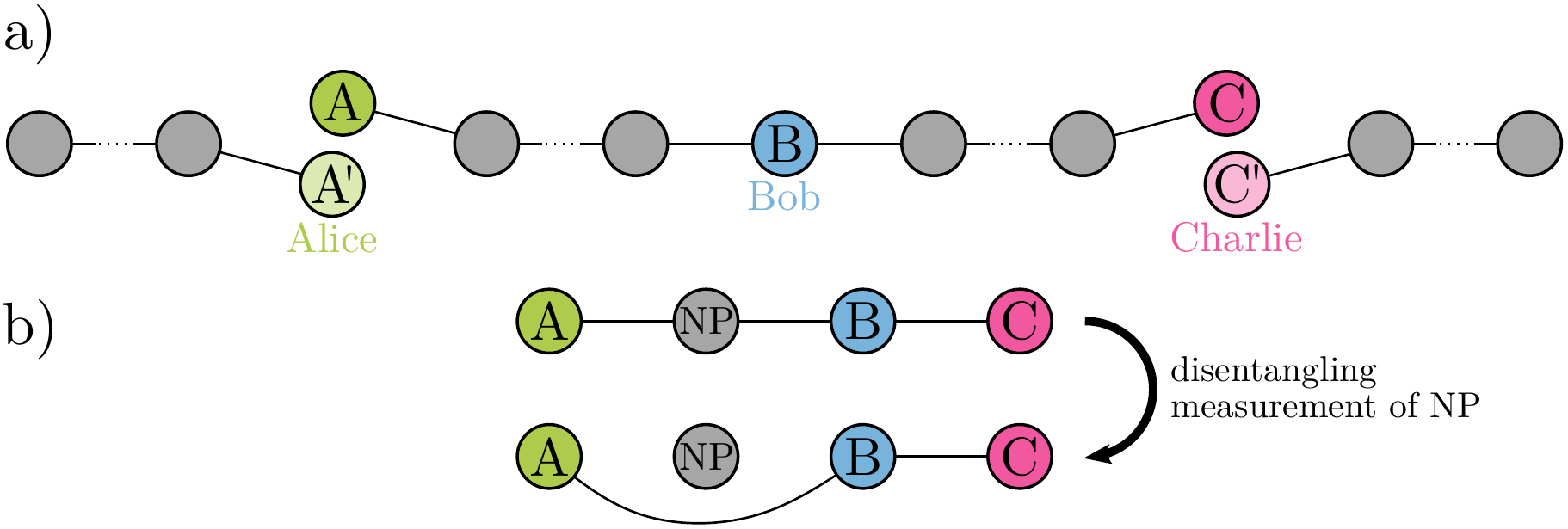}
		\caption{
			Sharing of the linear cluster state in the network. \textbf{a)} 
			The general protocol uses three linear cluster states, where Alice and Charlie have the first and last qubit, respectively, of the central linear cluster state, which is used for key generation. The use of three cluster states ensures the anonymity of all parties (see \cite{de2022anonymous} for details). \textbf{b)} In this work, we generate a 4-qubit linear cluster state acting as the central cluster state and extract a 3-qubit linear cluster state from it. 
		}
		\label{fig:networkduringprotocol}
	\end{centering}
\end{figure}

In photonic settings, like the one studied here, one often starts with two-qubit entangled states that are then fused to a larger cluster state~\cite{browne2005resource}. This cluster state is then shared in the network such that each party receives one qubit.

The general protocol for a network with $n$ parties has been introduced in~\cite{de2022anonymous}. Here, we focus on a network with four parties and aim at exchanging a key between three of the parties: Alice, Bob and Charlie (ABC). 
The protocol now works as follows: We generate a 4-qubit linear cluster state that is shared within the network.
The party not participating in the key exchange (NP) then performs a measurement. This measurement effectively removes their qubit from the 4-qubit cluster state and leaves a 3-qubit cluster state with ABC (see Fig.~\ref{fig:networkduringprotocol}). 

ABC measure their qubits in order to generate a key, exploiting the correlations of the 3-qubit linear cluster state which is locally equivalent to a 3-qubit GHZ state. This measurement type is called \textit{key generation}. As the exact 3-qubit state between ABC depends on the outcome of the disentangling measurement of the non-participating party, ABC perform bitflips conditioned on that measurement outcome. By measuring a stabilizer of the linear cluster state, ABC perform \textit{verification} measurements in order to detect possible eavesdropping attacks. The measurement bases are given in Appendix~\ref{sec:measurement_settings}. 

The protocol not only enables the generation of a secret key between ABC, but also keeps their identities secret.
Note that the parties are assumed to be honest but curious, i.e. they follow the protocol but if there is a way of gaining knowledge about the roles of other parties they will try to do this. If they follow the protocol applied in this work, however, they cannot learn anything about the positions of ABC in the network (apart from what is obvious from the network architecture), not even how many of them are sitting on the left or right of them~\cite{de2022anonymous}.
The non participating parties cannot learn anything about the identities of the other parties, even if they deviate from the protocol, as long as they do not collude.

We perform verification measurements in a percentage $p$ of the $L$ total rounds.
For both the key generation and the verification setting, we define the fraction of rounds with the incorrect outcome as $Q_{\text{keygen}}$ and $Q_{\text{verif}}$, respectively. The success rates of the key generation and the verification rounds are then given by $1-Q_{\text{keygen}}$ and $1-Q_{\text{verif}}$, respectively.
The parameter $Q_{\text{keygen}}$ is an upper bound for the pairwise bit error rates $Q_\text{keygen}^{A,B}$, between Alice and Bob, and $Q_\text{keygen}^{A,C}$, between Alice and Charlie. $Q_{\text{verif}}$ allows us to infer the maximal knowledge that a potential eavesdropper could gain about the key. In the post-processing steps, error correction and privacy amplification can be applied to the raw key to receive a correct and secret key. 
The number of key bits needed for those post-processing routines depends on the maximum of the pairwise bit error rates $Q_\text{keygen}^{A,B}$ and $Q_\text{keygen}^{A,C}$ for error correction and on $Q_{\text{verif}}$ for privacy amplification as well as on the desired level of security $\varepsilon_S$ (see Section 3.2 of \cite{de2022anonymous}).

%
%
\section{Experiment and Results}\label{sec:setup}
\begin{figure}[!ht]
	\begin{centering}
		\includegraphics[width=.45\textwidth]{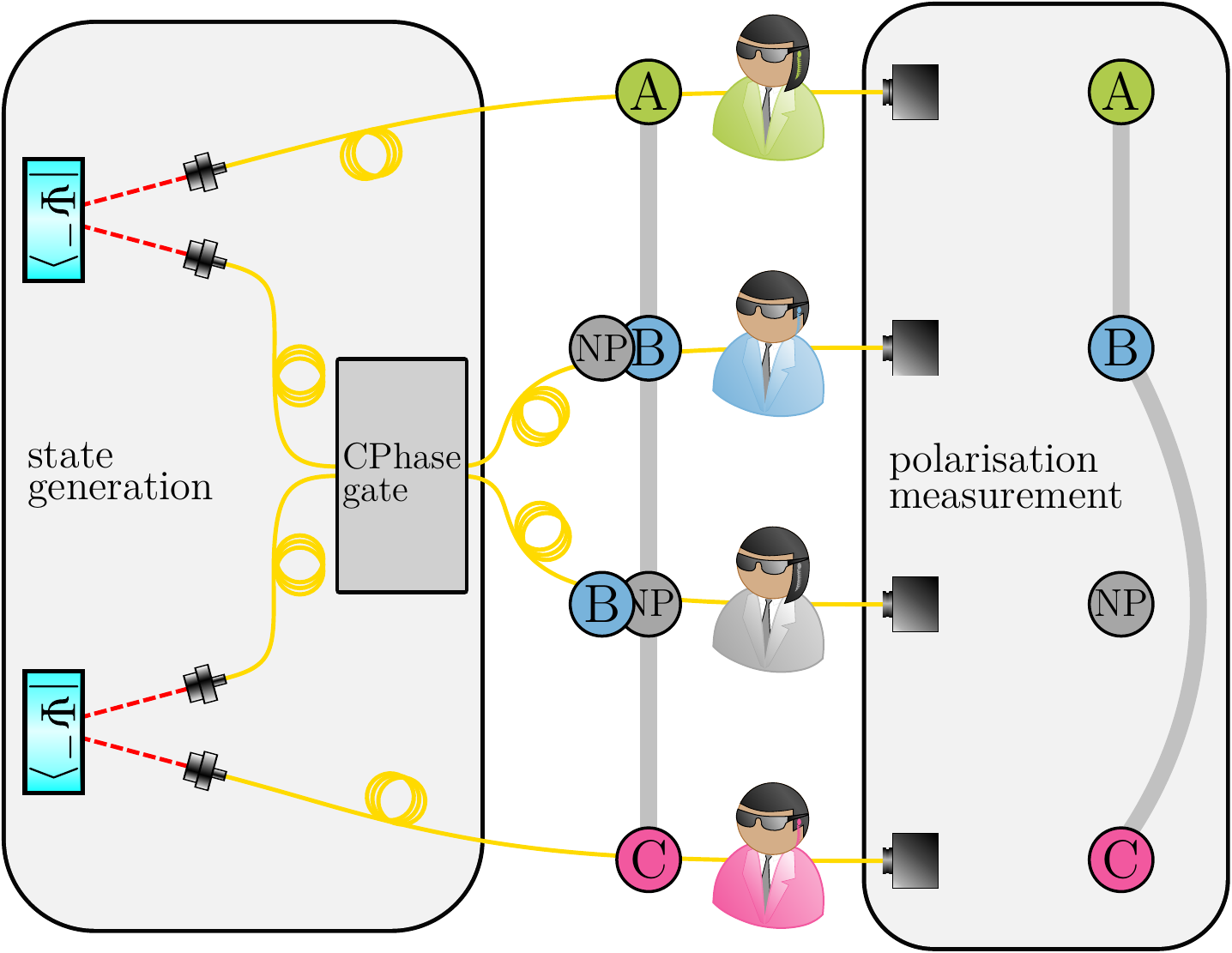}
		\caption{Schematic view of the implementation of the protocol. We generate two pairs of entangled photons in the state $\ket{\Psi^-}=\nicefrac{1}{\sqrt{2}}(\ket{H}\ket{V} - \ket{V}\ket{H})$. Applying a CPhase gate to one photon from each pair generates a four-photon cluster state upon postselection.  
		The four qubits are allocated to the different parties Alice, Bob, Charlie and a non-participating party. Polarisation measurements of the qubits can be realised in any basis and thus allow disentangling, key generation and verification operations.
			}
		\label{Fig:setup}
	\end{centering}
\end{figure}

In our implementation we generate two pairs of entangled photons and fuse them to a four-photon linear cluster state by applying a photonic CPhase gate \cite{kiesel2005} on one photon from each pair. 
We use polarisation encoding $\ket{0/1}=\ket{H/V}$. A scheme of the protocol implementation is shown and described in Fig.~\ref{Fig:setup}. The experimental setup is shown in Fig.~\ref{Fig:Detailedsetup}.

The generated state is characterised using quantum state tomography and a maximum likelihood estimation~\cite{james2001state_tomo}. The fidelity at a pump power of 400\,mW is estimated to $79.9 \pm 0.8$\,\%. The main source of noise is higher-order emission of the SPDC sources, which accounts for $4\,\%$ of all events. In addition, partial distinguishability as well as spectral mixedness affect the two photon interference at the polarisation-dependent beam splitter (PDBS) in the implementation of the CPhase gate.

\begin{figure}[b!]
	\begin{centering}
		\includegraphics[width=.45\textwidth]{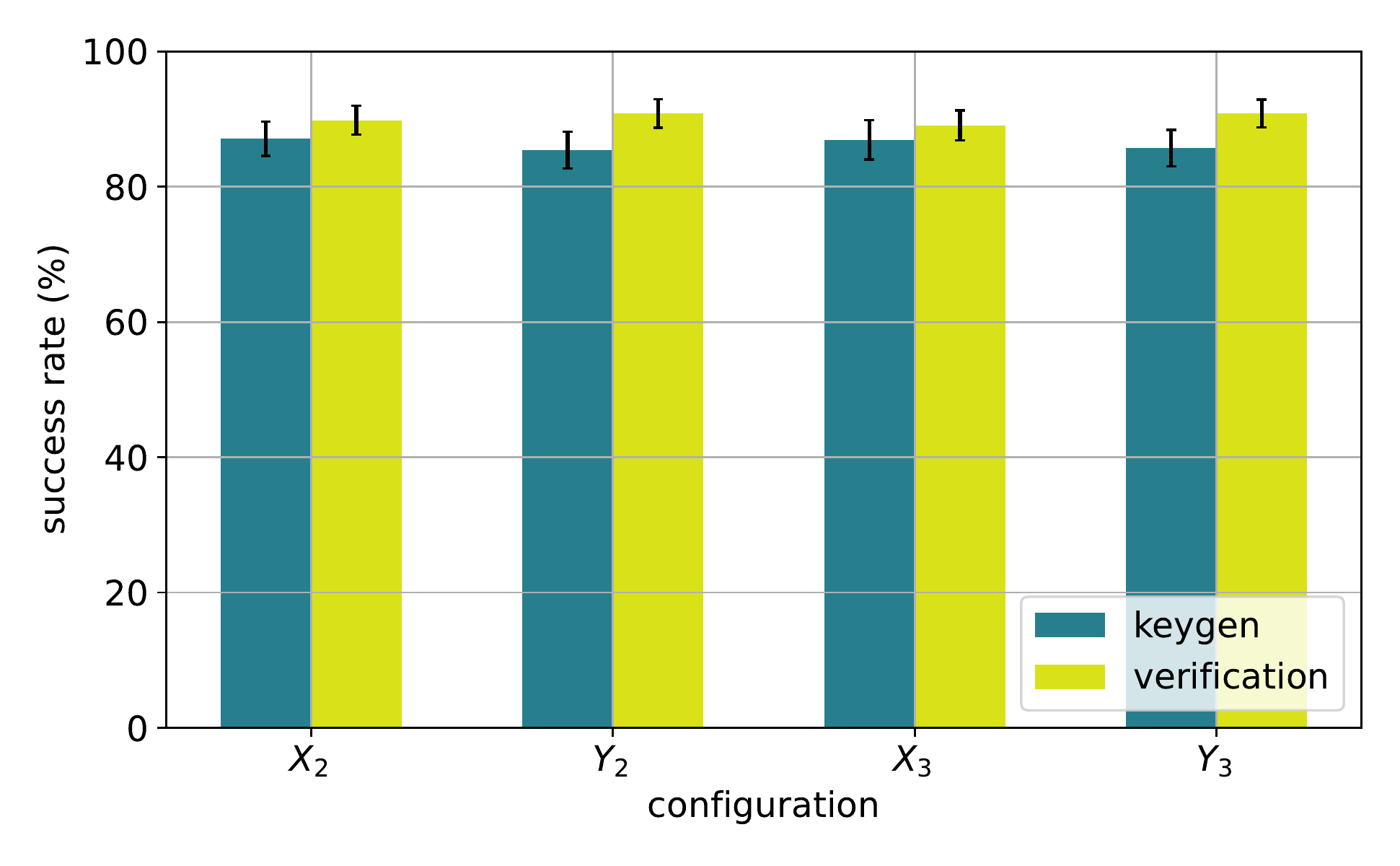}
		\caption{Success rate of key generation and verification rounds for the four different configurations.
		}
		\label{Fig:results}
	\end{centering}
\end{figure}

In our implementation, Alice and Charlie hold the first and last qubit of the state, respectively, meaning that either the party holding the second or third qubit is not participating.
In total there exist four experimental configurations, as either measurements in the $\sigma_X$- or $\sigma_Y$-basis can be used to disentangle the parties that do not participate while preserving entanglement between the participants.
We label those configurations $X_2$, $Y_2$, $X_3$, $Y_3$, where the letter indicates the type of disentangling measurement and the number the party removing themselves from the network.
The choice of the configuration determines the measurement settings of each party for the key generation and verification rounds.
We measure all eight measurement settings --- one key generation setting and one verification setting for each of the four configurations --- and determine the success rate for each setting (see Fig.~\ref{Fig:results}). 

For the implementation of the protocol in a realistic setting, we choose randomly whether key generation or verification is performed using a biased random number generator.
A single fourfold event is considered a round. For each setting we integrate over a time of 60\,s, which we call a run containing multiple rounds. At the start of each run, the biased random number generator indicates if the next run is a key generation run or a verification run.


We set $p=10\,\%$ and run the protocol; we exchange 41033 bits and use those to encrypt a binary image. In addition, we measure 3794 rounds for verification. 
From the measurements we obtain a success probability of $(87.76 \pm 0.54)\,\%$ for the key generation rounds and  $(87.01 \pm 0.55)\,\%$ for the verification rounds.
Note that because the random number generator is called only a finite number of times, the ratio of verification rounds from total rounds slightly differs from the value of $p$.

\begin{figure*}[t!]
	\begin{centering}
		\includegraphics[width=\textwidth]{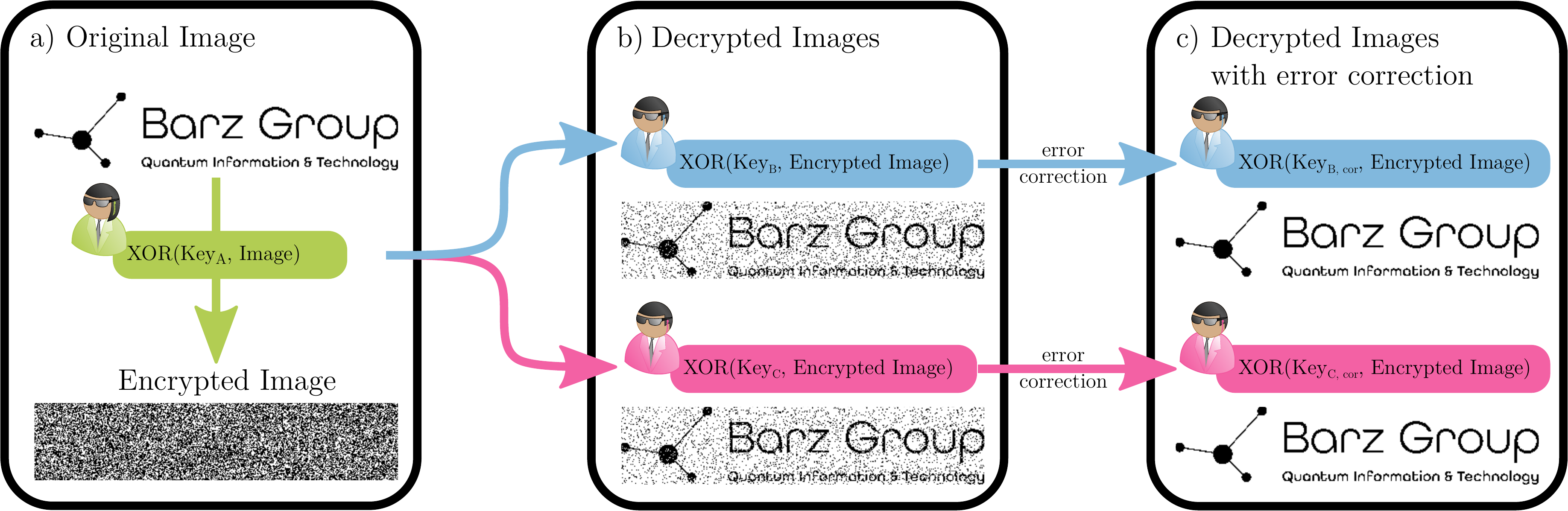}
		\caption{Example for an encryption and decryption with the generated key. \textbf{a)} Alice encrypts a binary image (top) using the key and sends the encrypted image (bottom) to Bob and Charlie. \textbf{b)} Bob and Charlie use their keys to decrypt the sent encrypted image. When using the erroneous raw key to do so, the decrypted image deviates from the original. \textbf{c)} After the application of the error correction method the original image is retrieved using the corrected keys $\text{Key}_\text{B, cor}$ and $\text{Key}_\text{C, cor}$ for the decryption.
		}
		\label{Fig:encrypt_decrypt}
	\end{centering}
\end{figure*}

We use the key to encrypt a binary image by performing an XOR operation for every image pixel with a bit from Alice's key (see Fig.~\ref{Fig:encrypt_decrypt}). 
Bob and Charlie can decrypt the image using their keys: 
If the bits of the encryption and decryption key are the same, the original image pixel is retrieved. However, if due to errors, bits of the encryption and decryption key are different, this will result in an incorrectly communicated image pixel. In our implementation, this noise arises from imperfections in the state preparation and transmission. As a result, classical post-processing is necessary because, with an erroneous key, the sent message will also contain errors.

\textbf{Error correction.} We use low-density parity check codes (LDPC) to perform \textit{error correction} (see Appendix~\ref{sec:app_err_corr}).
First, ABC calculate parity check bits from the raw key. For this, they use a publicly known sparse matrix which indicates the raw key bits that have to be binary added in order to determine each parity bit. Then, Alice sends her parity bits to Bob and Charlie via a classical channel. 
Comparing Alice's parity check bits with those obtained from their keys allows them to estimate which bits were (most likely) subject to noise and therefore flipped. The identified bits are then corrected by flipping them back. 

The ratio $r$ between the number of raw key bits and the number of parity bits is called the \textit{code rate} and its chosen value is dependent on the error rate. 
If $r$ is too high, meaning if not enough parity bits are used, not all errors can be corrected.
For different values of $r$ and depending on $Q_\text{keygen}^{A,B}$ and $Q_\text{keygen}^{A,C}$, respectively, the error can be corrected partially or completely (see Tab. \ref{tab:error_corr}). For the keys in this work, all errors could be corrected using a code rate of $r = 0.5$. A detailed explanation and the chosen parameters of the error correction procedure can be found in the Appendix~\ref{sec:app_err_corr}.


\begin{table}[b]
	\caption{Ratio of bits different from Alice's key in the keys of Bob and Charlie for the raw key and error corrected keys with different code rates $r$. 
	}\label{tab:error_corr}
	\begin{tabular*}{0.9\linewidth}{c @{\extracolsep{\fill}} rr}
		\toprule
		& Bob (\%) & Charlie  (\%)\\
		\midrule
		raw key & 10.37 & 9.67 \\
		$r=2:3$ & 10.22 & 8.88 \\		
		$r=3:5$ & 6.82 & 0 \\
		$r=1:2$ & 0 & 0 \\
		\bottomrule
	\end{tabular*}
\end{table}

%
%

\section{Analysis of the protocol under realistic conditions}\label{sec:analysis}

In a realistic setting, errors in the key could be introduced by an eavesdropper trying to gain knowledge about the key, thus compromising the security of the protocol. Therefore, privacy amplification is needed in addition to error correction.
For both error correction and privacy amplification, a fraction of the exchanged key has to be used.
The fraction of the key needed for those post-processing steps should be small compared to the key length in order to obtain a positive key rate. It depends on 
the parameters $Q_{\text{verif}}$ and $\max(Q_\text{keygen}^{A,B},Q_\text{keygen}^{A,C})$ as they indicate the level of information leakage and errors in the key, respectively. If $Q_{\text{verif}}$ and $\max(Q_\text{keygen}^{A,B},Q_\text{keygen}^{A,C})$ are too high, no secure and correct key can be achieved using the post-processing steps.

The \textit{asymptotic key rate} (AKR) is an upper bound to the maximal achievable fraction of the raw key which can be used as a correct and secure key. For the protocol used here, it is given by

\begin{equation}\label{eq:akr}
	\mbox{AKR} = 1 - h\left(Q_{\text{verif}}\right) - h\left(\max{\left(Q_\text{keygen}^{A,B},Q_\text{keygen}^{A,C}\right)}\right),
\end{equation}

where $h$ is the binary entropy $h(x):= - x\log_2(x) - (1-x)\log_2(1-x)$. The binary entropy takes values between $0$ and $1$, therefore the AKR has a value between $-1$ and $1$. A negative value indicates that $Q_{\text{verif}}$ and $\max{(Q_\text{keygen}^{A,B},Q_\text{keygen}^{A,C})}$ are too large and thus the post-processing steps cannot be carried out.

\begin{figure}[t!]
	\begin{centering}
		\includegraphics[width=.45\textwidth]{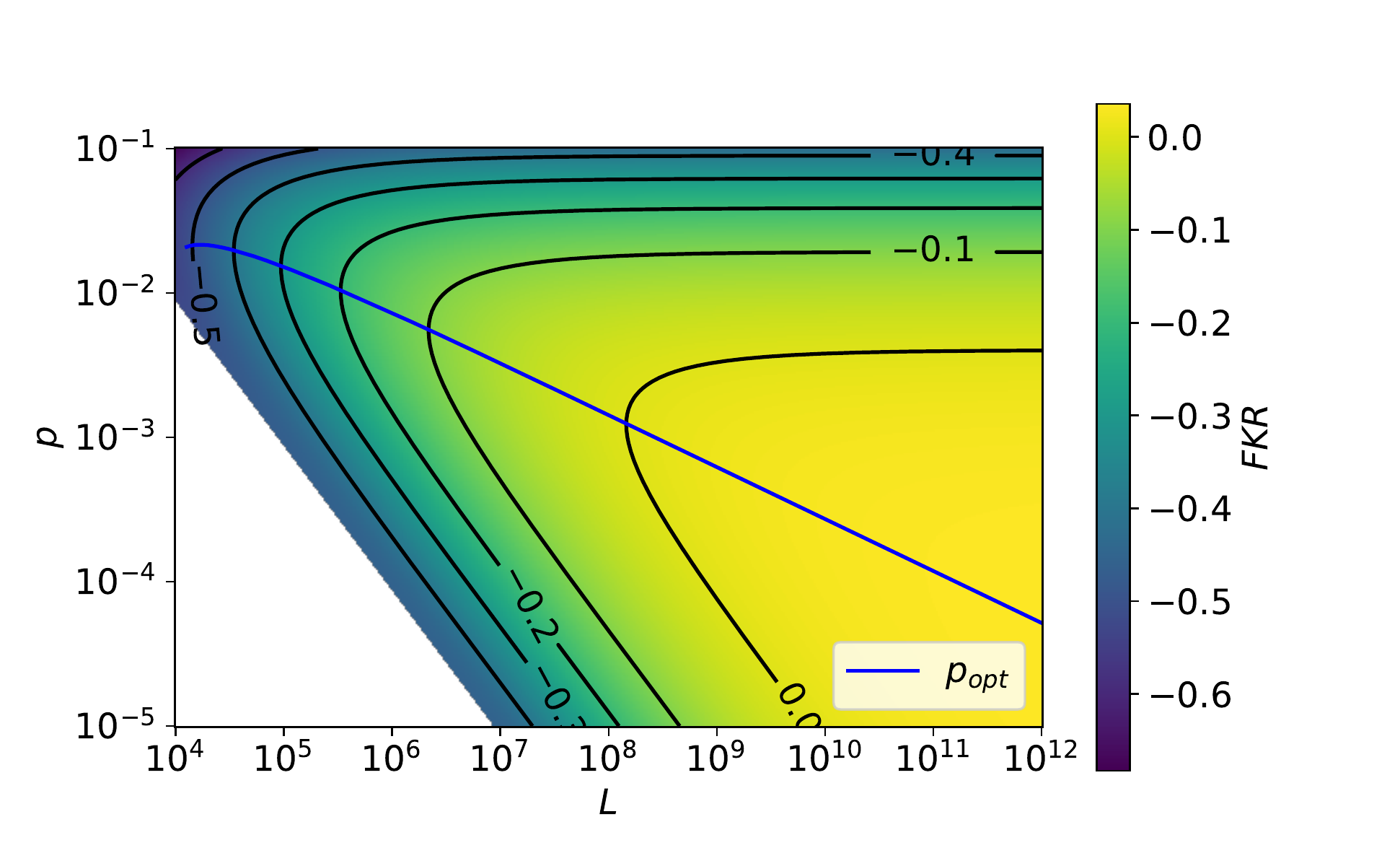}
		\caption{Dependence of finite key rate FKR on the raw key length $L$ and $p$. The minimal $L$ at which one can get a positive FKR with the parameters estimated in this setup is given by $L = 1.46\cdot10^8$ when choosing a $p = 0.12\,\%$. The blue line indicates the optimal choice of $p$ for each $L$, meaning the value of $p$ at which the FKR becomes maximal for each fixed $L$. The optimal $p$, meaning the $p$ with which the highest FKR is achieved, becomes smaller the higher $L$ gets. In the limit of $L \to \infty$, the optimal value of $p$ goes to zero. Therefore, the AKR does not depend on $p$.
		}
		\label{Fig:FKR}
	\end{centering}
\end{figure}

For finite keys of length $L$ the communication of the verification rounds and statistical uncertainties in the estimation of the errors lead to a smaller ratio of secure and correct key to raw key. This ratio is called the \textit{finite key rate} (FKR). The AKR is the limit of the FKR for $L \to \infty$.
Fig.~\ref{Fig:FKR} shows how the choice of the parameters $p$ and $L$ affect the FKR for the values of $Q_{\text{verif}}$ and $\max{(Q_{\text{keygen}}^{A,B},Q_\text{keygen}^{A,C})}$ estimated in our setup.

We perform another run of the protocol in the configuration $X_2$ and set $p = 2\,\%$. We retrieved 11108 rounds in total, including 10814 key generation rounds and 294 verification rounds. The estimated values of $Q_{\text{verif}} = (11.2 \pm 1.8)\,\%$ and $\max(Q_\text{keygen}^{A,B},Q_\text{keygen}^{A,C}) = (9.59 \pm 0.28)\,\%$
correspond to $h(Q_{\text{verif}}) = 0.507 \pm 0.055$ and $h(\max(Q_{\text{keygen}}^{A,B},Q_\text{keygen}^{A,C})) = 0.456 \pm 0.001$, thus leading to a positive value of $\mbox{AKR} = 0.0375 \pm 0.0557$.

%
%
\section{Conclusion and Outlook}\label{sec:conclusion}

In this work, we demonstrate anonymous quantum conference key agreement using a linear cluster state. We showed the encryption of a picture and analysed the security parameters in the experimental setup and from this the finite and (positive) asymptotic key rate. The change of the network configuration --- meaning who belongs to the communicating subgroup and who does not --- only requires the change of the measurement settings, which makes the introduced protocol a feasible and flexible technique of networked communication. This work widens the range of possible resource states used for such networked communication tasks. As cluster states are an important resource in quantum computation, this could open an avenue to the hybrid use of these resource states for networked computing and communication.

It will be interesting to extend the given network to larger states and networks. When going to such larger systems, questions for an efficient usage of the states arise: Is it possible to use one resource state for several communicating parties in parallel? Can parts of a state still be used if the state was due to losses not entirely transmitted? Furthermore, a detailed study of the noise occurring in the implementations will be the key to developing adapted protocols.

A challenging task on the way to such larger networks remains the creation of multipartite resource states. New techniques in the state generation, such as the use of quantum dots as on-demand sources \cite{uppu2021quantum}, could provide a new avenue to the generation of cluster states and their use in the context of secure key exchange.

\section{Author's contributions}

L.R. and J.B. set up the experiment, measured the data for the implementation of the protocols, and analysed the data. J.d.J., F.H.  and A.P. performed the theoretical analysis, S.B. led the project. All authors discussed the results and contributed to writing and commenting on the manuscript.

\begin{acknowledgments}
We thank Nathan Walk and Jens Eisert for useful discussions, and Christopher Thalacker for setting up the early stages of the experiment.
L.R., J.B. and S.B. acknowledge support from the Carl Zeiss Foundation, the Centre for Integrated Quantum Science and Technology (IQ$^\text{ST}$), the German Research Foundation (DFG), the Federal Ministry of Education and Research (BMBF, project SiSiQ and PhotonQ), and the Federal Ministry for Economic Affairs and Energy (BMWi, project PlanQK). J.d.J. and A.P. acknowledge support from the Emmy Noether DFG grant No. 418294583. F.H. acknowledges support from the German Academic Scholarship Foundation. A.P.  also acknowledges support from the Einstein Research Unit on Quantum Devices.
\end{acknowledgments}


%

\newpage

%
%

\onecolumngrid

\appendix

\section{Setup}

\begin{figure}[ht!]
	\centering
	\includegraphics[width=.99\textwidth]{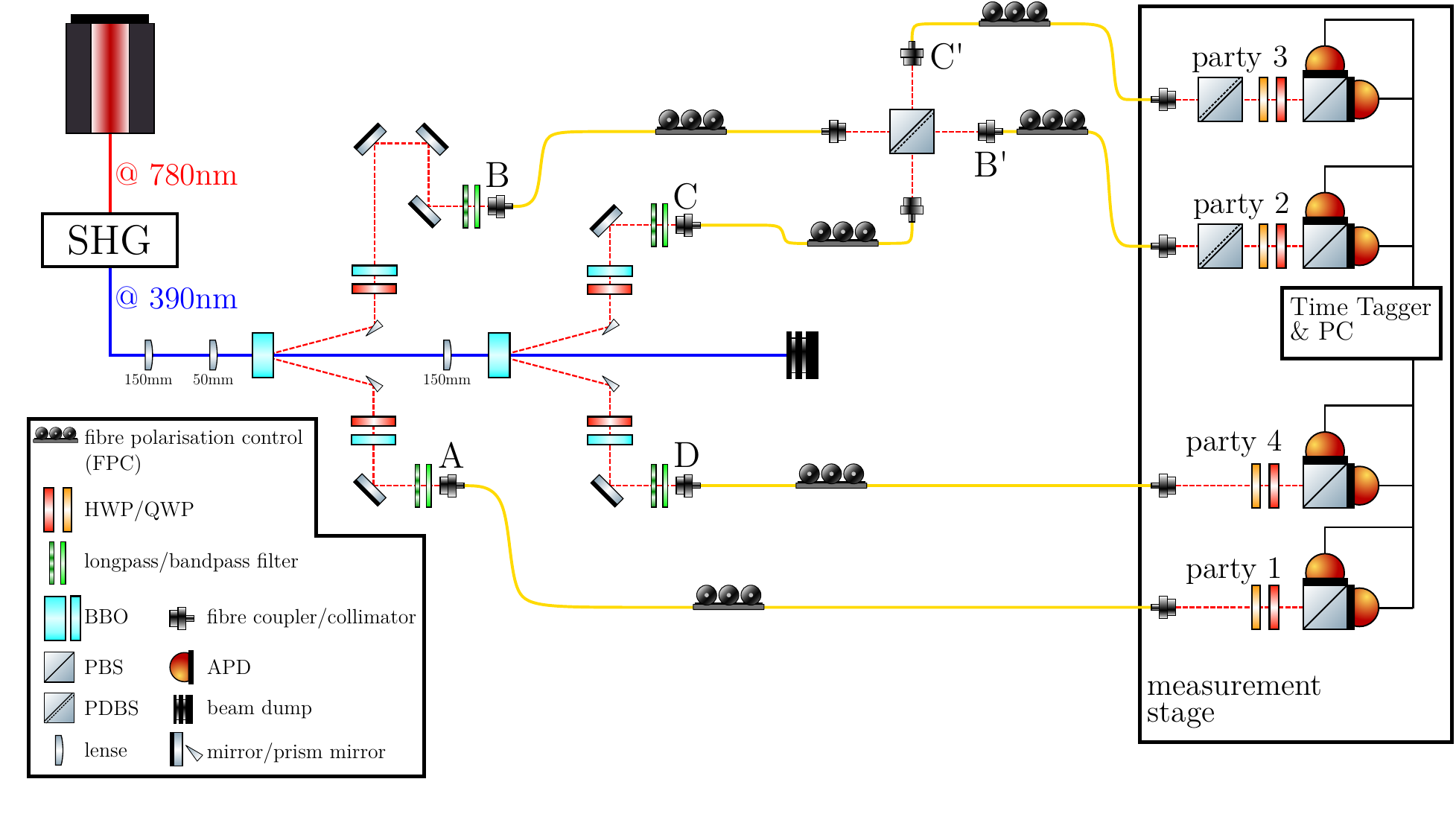}
	\caption{Experimental setup. We generate two entangled photon pairs by pumping two Barium Borate (BBO) crystals cut for type-II-spontaneous parametric down conversion (SPDC) with a pulsed Titanium-Sapphire laser ($\tau=140\,fs$, $\lambda=780\,nm$ upconverted to $\lambda=390\,nm$, $P=400\,mW$).
		The generated Bell pairs are fibre coupled and one photon of each Bell pair is overlapped at a polarisation-dependent beam splitter (PDBS).
		This PDBS transmits horizontally polarised photons and reflects 2/3 of the vertically polarised photons. Two further PBDS, one in each output mode of the first, reflect 2/3 of the $\ket{H}$ photons and transmit all $\ket{V}$ photons.
		All PDBS together form a photonic CPhase gate \cite{kiesel2005} upon coincidence detection, postselecting all events which results in one photon per output mode (1/9 of the cases).
		For each channel a combination of a half wave plate (HWP), a quarter wave plate (QWP) and polarizing beam splitter (PBS) enables the measurement of the respective qubit in the necessary bases, followed by an avalanche photo diode for photon detection. 
		The setup generates a state which is locally equivalent to a four-photon linear cluster state up to local unitaries which we absorb in the measurement bases.
	}
	\label{Fig:Detailedsetup}
\end{figure}

\FloatBarrier

\newpage

\section{Measurement settings of the protocol}\label{sec:measurement_settings}

In our setup, we generated the state
$\ket{LC'_4} = \frac{1}{2} \left( \ket{HVHV} + \ket{HVVH} - \ket{VHHV} + \ket{VHVH} \right)$, which is locally equivalent to the state $\ket{LC_4} := \mbox{CZ}_{1,2}\mbox{CZ}_{2,3}\mbox{CZ}_{3,4}\ket{\mbox{++++}} = \frac{1}{2} \left( \ket{+00+} + \ket{+01-} +\ket{-10+} -\ket{-11-} \right)$. The two states are related by $\ket{LC'_4} = H_1 X_2 X_3 H_4 \ket{LC_4}$. The $H$ denotes a Hadamard gate, the $X$ a Pauli-$X$-gate and the indices indicate on which qubit the gates act. The measurement bases for the different configurations are listed in Tab.~\ref{tab:measurement_settings}.

\begin{table}[h!]
	\caption{Measurement settings for the different network configurations. The indices $2/3$ at the configuration denote which party is the non participating one and the capital letter in which basis $\sigma_X/\sigma_Y$ the measurement is applied. For each configuration the measurement setting for each party 1-4 is shown for key generation and verification purposes. $Z$ denotes a measurement in the $\sigma_Z$ basis. The measurements are realised by setting the quarter wave plate and half wave plate to the corresponding angles. With the notation (angle QWP, angle HWP), these are ($45^\circ$, $22.5^\circ$) for $X$, ($45^\circ$, $0^\circ$) for $Y$ and ($0^\circ$, $0^\circ$) for $Z$.}\label{tab:measurement_settings}
	\begin{tabular*}{0.8\linewidth}{c @{\extracolsep{\fill}} cc}
		\toprule
		configuration & key generation & verification \\
		\midrule
		$X_2$ & $X_1 X_2 Z_3 Z_4$ & $Z_1 X_2 X_3 X_4$ \\
		$Y_2$ & $Y_1 Y_2 Z_3 Z_4$ & $Z_1 Y_2 X_3 X_4$ \\
		$X_3$ & $Z_1 Z_2 X_3 X_4$ & $X_1 X_2 X_3 Z_4$ \\
		$Y_3$ & $Z_1 Z_2 Y_3 Y_4$ & $X_1 X_2 Y_3 Z_4$ \\
		\bottomrule	
	\end{tabular*}
\end{table}

\FloatBarrier

\section{Error correction procedure}\label{sec:app_err_corr}
In order to correct the errors $\text{Err}(\text{Key}_A,\text{Key}_B)$ and $\text{Err}(\text{Key}_A,\text{Key}_C)$ between the key of Alice and the keys of the other participants, low density parity check (LDPC) matrices are used. 
Specifically the DVB-S2 standard \cite{Morello2006} is applied providing matrices of the form 
\begin{equation}
	H = [H'|S]
\end{equation}
where $H'$ is a sparse matrix of dimension $(N-k)\times k$ and $S$ is a staircase matrix of dimension $(N-k)\times(N-k)$. Possible values of $N$ within the standard are 64800 and 16200. Due to the length of the created key the latter one is chosen. Depending on the error rate a code rate of $r=k/N$ is chosen. A higher $r$ hereby corresponds to less parity check bits and hence is used for low error rates. In order to correct all errors in Bobs and Charlie's keys, $r$ is set to $1/2$. 

Alice divides her key in blocks of $k$ bits and calculates the parity check bits for each block using $H$. Bob and Charlie receive these parity check bits over a classical verified channel. At Bobs and Charlie's site they have now their respective key containing errors and the check bits of Alice, which are assumed to be transmitted without errors. With the knowledge of $H$, Bob and Charlie can correct their errors.
For this, the \texttt{ldpcEncode} and \texttt{ldpcDecode} functions provided by MATLAB are used. For the latter the belief propagation algorithm is chosen \cite{Gallager1963}.

\end{document}